\documentclass[a4paper]{article}

\usepackage{INTERSPEECH2020}
\usepackage{scalerel}
\usepackage{multirow}
\usepackage{cite}

\def\Z{{\cal Z}}
\def\vx{\mathbf{x}}
\def\vh{\mathbf{h}}
\def\vy{\mathbf{y}}

\def\sy{y}
\def\hy{\hat{y}}

\def\A{{\cal A}}

\def\blank{{\left<\texttt{b}\right>}}

\title{Conv-Transformer Transducer: Low Latency, Low Frame Rate, Streamable End-to-End Speech Recognition}
\name{Wenyong Huang, Wenchao Hu, Yu Ting Yeung, Xiao Chen}
\address{
  Huawei Noah's Ark Lab}
\email{\{wenyong.huang, huwenchao, yeung.yu.ting, chen.xiao2\}@huawei.com}

\begin{document}

\maketitle

\begin{abstract}
Transformer has achieved competitive performance against state-of-the-art end-to-end models in automatic speech recognition (ASR), and requires significantly less training time than RNN-based models.
The original Transformer, with encoder-decoder architecture, is only suitable for offline ASR.
It relies on an attention mechanism to learn alignments, and encodes input audio bidirectionally.
The high computation cost of Transformer decoding also limits its use in production streaming systems.
To make Transformer suitable for streaming ASR, we explore Transducer framework as a streamable way to learn alignments.
For audio encoding, we apply unidirectional Transformer with interleaved convolution layers.
The interleaved convolution layers are used for modeling future context which is important to performance.
To reduce computation cost, we gradually downsample acoustic input, also with the interleaved convolution layers.
Moreover, we limit the length of history context in self-attention to maintain constant computation cost for each decoding step.
We show that this architecture, named Conv-Transformer Transducer, achieves competitive performance on LibriSpeech dataset (3.6\% WER on test-clean) without external language models.
The performance is comparable to previously published streamable Transformer Transducer and strong hybrid streaming ASR systems, and is achieved with smaller look-ahead window (140~ms), fewer parameters and lower frame rate.

\end{abstract}

\section{Introduction}

Transformer \cite{vaswani2017attention} models have achieved state-of-the-art results in many natural language processing tasks \cite{devlin2019bert,dai2019transformer}.
Recently, Transformer is gaining popularity in speech recognition research community \cite{dong2018speech,karita2019improving,zeyer2019comparison,karita2019comparative}.
Transformer-based speech recognition models have achieved comparable or better performance than state-of-the-art models, which are mostly based on recurrent neural network (RNN).
Transformer is more parallelizable than RNN, thus is significantly faster to train \cite{vaswani2017attention,karita2019comparative}.

Despite its success, the original Transformer with encoder-decoder architecture is only suitable for offline speech recognition.
There are a number of challenges to apply the original Transformer for streaming speech recognition. First, the original Transformer relies on an attention mechanism over full encoder output to learn alignments between input and output sequences \cite{bahdanau2014neural}. Second, the original Transformer encodes input audio in a bidirectional way, thus requires a full utterance as input. Moreover, computation of Transformer increases in quadratic complexity with the length of input sequences. Naive implementation usually leads to much slower decoding than RNN-based models \cite{karita2019comparative}.

For alignment learning, there exist many streamable ways for speech recognition, e.g., Connectionist Temporal Classification (CTC) \cite{graves2006connectionist}, Transducer \cite{graves2012sequence}, Monotonic Chunkwise Attention (MoChA) \cite{chiu2018monotonic}, Triggered Attention \cite{moritz2019triggered}.
All of these alignment learning methods can be combined with Transformer. In this work, we focus on Transducer, as previous works \cite{he2019streaming,sainath2020streaming} suggest that Transducer models outperform traditional hybrid models for streaming speech recognition in production settings.
Several research groups have been working on combining Transformer with Transducer for speech recognition, which is usually referred to as Transformer Transducer \cite{tian2019self,yeh2019transformer,zhang2020transformer}.

To make Transformer encoder streamable, previous works of Transformer Transducer \cite{yeh2019transformer,zhang2020transformer} tried to limit future context (right-context) in self-attention for input audio encoding.
Although each layer only requires a little future context, the overall future context aggregates over all Transformer layers. The overall look-ahead window is still large.
In this work, we explore using unidirectional Transformer with interleaved convolution layers for audio encoding.
Unidirectional Transformer requires no future context. Thus it is streamable and does not introduce latency into the model.
However, future context is important to speech recognition performance. Therefore, we add interleaved convolutions between Transformer layers to model future context. We carefully control number of layers and filter size of convolution layers. In our model, the look-ahead window introduced by convolution layers is 140~ms.

We explore a number of options for practical streaming speech recognition with Transformer. First, we apply interleaved convolutions to gradually downsample input audio sequence. This enables our model to run in a low frame rate of 80~ms, thus reduces computation cost significantly \cite{pundak2016lower}. Second, we limit the length of history context (left-context) of self-attention in Transformer layers to maintain constant computation cost for each decoding step. Finally, we apply relative position encoding \cite{dai2019transformer}, which enables hidden state reuse for Transformer. Hidden state reuse improves decoding speed of Transformer significantly \cite{dai2019transformer}.

We name our design as Conv-Transformer Transducer. We show that this architecture achieves competitive performance on LibriSpeech dataset \cite{panayotov2015librispeech}, with word error rate (WER) of 3.5\% in test-clean and 8.3\% in test-other without external language models. When we further limit history context of self-attention, our proposed model still achieves WER of 3.6\% and 8.9\% respectively. The performance is comparable to previously published streamable Transformer-Transducer \cite{zhang2020transformer}, and strong hybrid speech recognition systems, with smaller look-ahead window, fewer parameters and lower frame rate.

This paper is organized as follows. In the next Section, we introduce the Transducer framework and detailed architecture of Conv-Transformer Transducer. In Section 3, we present our experimental results on LibriSpeech dataset. Finally, we conclude our work in Section 4.

\section{Conv-Transformer Transducer}
\subsection{Transducer}

Transducer, proposed in \cite{graves2012sequence}, is a neural sequence transduction framework which can be used for end-to-end speech recognition, i.e., transduction of an input acoustic sequence into an output transcription label sequence.

\begin{figure}[t]
  \centering
  \includegraphics[width=0.5\linewidth]{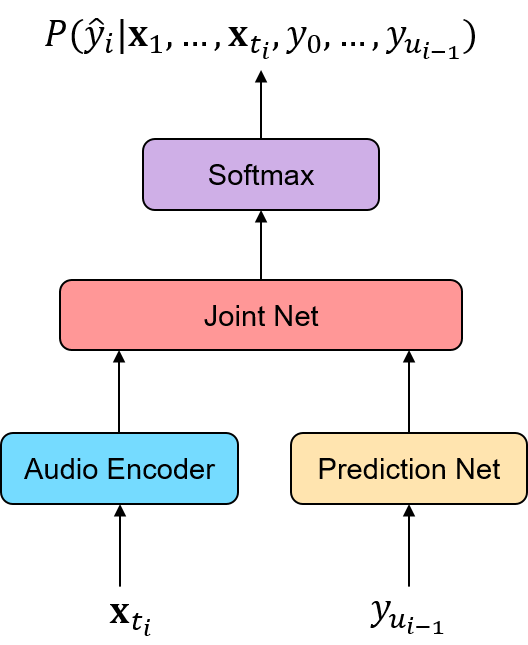}
  \caption{Transducer model architecture.}
  \label{fig:transducer}
  %\vspace{-2mm}
\end{figure}

We denote an input acoustic sequence with $T$ frames as $\vx = (\vx_1, \ldots, \vx_T)$, and a transcription label sequence of length $U$ as $\vy = (\sy_1, \ldots, \sy_{U})$, where $\sy_u \in \Z$ and $\Z$ is vocabulary of output labels.
As depicted in Figure~\ref{fig:transducer}, a Transducer model first encodes an input acoustic sequence with an audio encoder network, producing a sequence of encoder states denoted as $\vh = (\vh_1, \ldots, \vh_T)$.
For each encoder state $\vh_t$, the model first predicts either a label or a special blank symbol $\blank$ with a joint net. When the model predicts a label, the model continues to predict the next output. When the model predicts a blank symbol, which indicates that no more labels can be predicted, the model proceeds to the next encoder state.
This process is similar to CTC \cite{graves2006connectionist}, but there is a difference.
A Transducer model makes use of previously predicted non-blank labels as input condition to predict the next output.
The previously predicted labels are encoded with another network referred to as prediction net.

During the process described above, the Transducer model defines a conditional distribution,
\begin{equation}
  % \small
P(\hat{\vy} | \vx) = \prod_{i=1}^{T+U} P(\hy_{i} | \vx_1, \cdots, \vx_{t_i}, y_0, \ldots, y_{u_{i-1}})
\label{eqn:transducer-path-prob}
\end{equation}
where $\hat{\vy} = (\hy_1, \ldots, \hy_{T+U}) \subset \{\Z\cup\blank\}^{T+U}$ is an alignment path with $T$ blank symbols and $U$ labels such that removing all blank symbols in $\hat{\vy}$ yields $\vy$, and $y_0$ represents start of sentence.
To obtain probability of target sequence, Transducer computes a marginalized distribution,
\begin{equation}
  \small
P(\vy | \vx) = \hspace{-0.23in} \sum_{\hy \in \A_{\text{ALIGN}}(\vx, \vy)} P(\hat{\vy} | \vx)
\label{eqn:align-rnnt}
\end{equation}
\noindent where $\A_{\scaleto{ALIGN}{3pt}}(\vx, \vy)$ is the set containing all valid alignment paths such that removing the blank symbols in $\hat{\vy}$ yields $\vy$.
The summation of probabilities of all alignment paths is computed efficiently with forward-backward algorithm \cite{graves2012sequence}.

In most of previous works in Transducer framework, audio encoder and prediction net are composed of RNN \cite{graves2012sequence}. Such type of Transducers are often referred to as Recurrent Neural Network Transducer (RNN-T).
Other types of networks are allowed in Transducer model.

\subsection{Architecture of Conv-Transformer Transducer}

\begin{figure}[t]
  \centering
  \includegraphics[width=\linewidth]{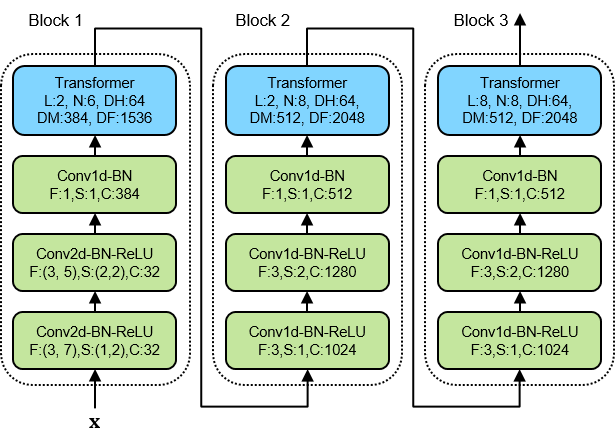}
  \caption{Audio encoder of Conv-Transformer Transducer: \textbf{ConvND-BN-RELU} denotes N-Dimensional convolution layers followed by batch normalization and Rectified Linear Unit (ReLU) activation. For convolution layers, \textbf{F} denotes number of filters, \textbf{S} denotes the value of stride, \textbf{C} denotes number of output channels. For 2D convolutions, the first values of \textbf{F} and \textbf{S} correspond to time dimension, the second values correspond to frequency dimension.
  For Transformer, \textbf{L} denotes number of layers, \textbf{N} denotes number of heads, \textbf{DH} denotes dimension of each head, \textbf{DM} denotes input and output dimension of self-attention layers and feed-forward network (FFN) layers, \textbf{DF} denotes dimension of intermediate hidden layer in the FFN.}
  \label{fig:conv_transformer_encoder}
  \vspace{-3mm}
\end{figure}

We now describe the architecture of our Conv-Transformer Transducer.
The audio encoder is illustrated in Figure~\ref{fig:conv_transformer_encoder}, which is composed of three blocks. Each block is composed of three convolution layers followed by unidirectional Transformer.
Unidirectional Transformer is similar to encoder of the original Transformer, except that self-attention is masked, preventing the self-attention from attending to subsequent positions beyond current position. Unidirectional Transformer does not require future context, and is widely used in language modeling \cite{al2019character,dai2019transformer} and text generation \cite{liu2018generating}.

\begin{figure}[t]
  \centering
  \includegraphics[width=1.0\columnwidth, trim={3.1cm 11.1cm 3.5cm 10.7cm},clip]{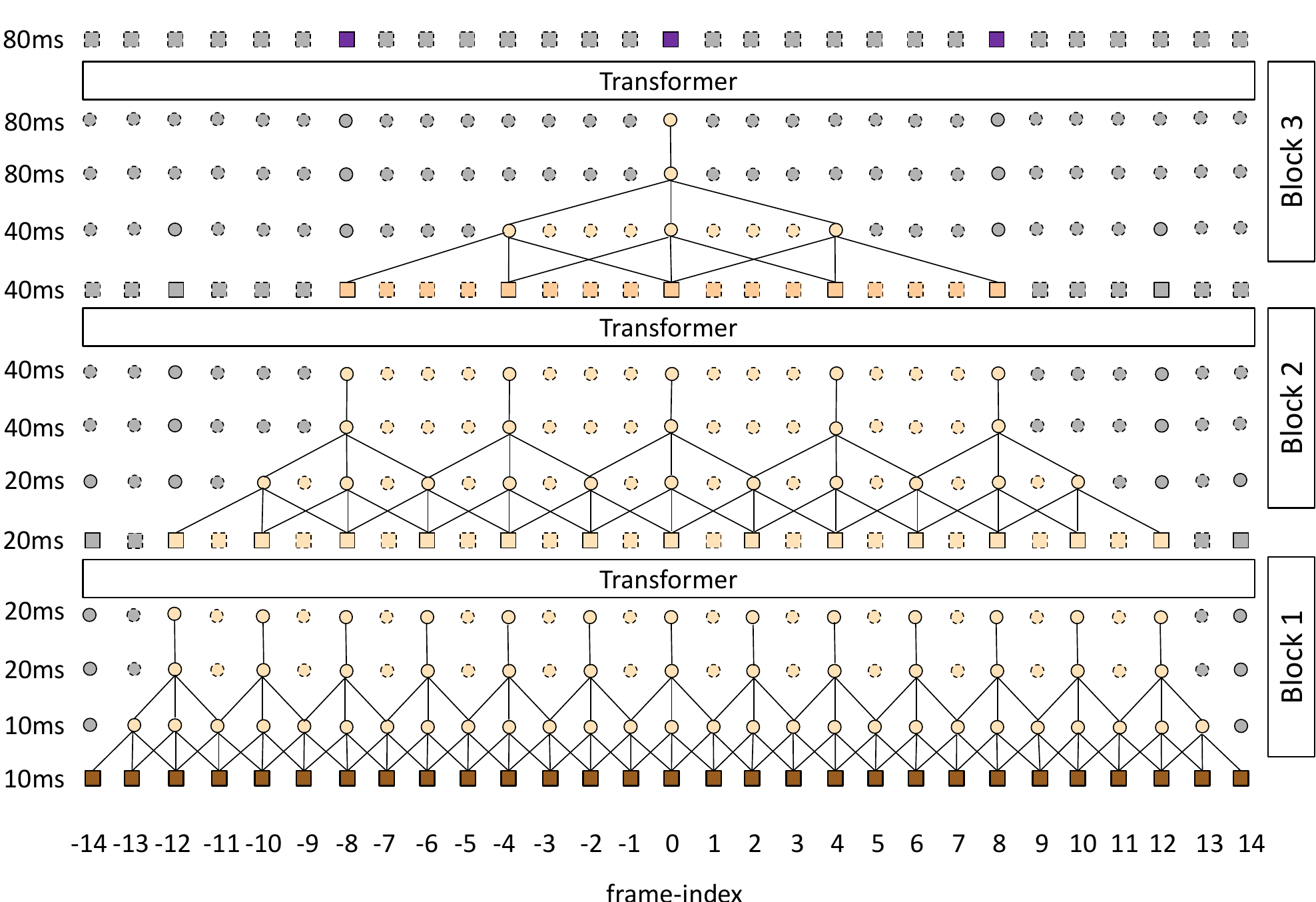}
  \caption{Illustration of context window and frame rate change of convolution layers in audio encoder of Conv-Transformer Transducer at current input frame (frame-index = 0). Input and output of each block are represented by squares. Convolution layers are represented by circles. Only time dimension is shown. For each layer, activation with dashed outline is skipped in computation.}
  \label{fig:delay-count}
    %\vspace{-0.2cm}
\end{figure}

However, future acoustic context is helpful to improve recognition accuracy.
Therefore, we apply convolution layers before Transformer in each block.
In our model, all future context comes from convolution layers. The size of look-ahead window required by current input frame is illustrated in Figure \ref{fig:delay-count}. In our 3-block setting, the size of look-ahead window is 140~ms, which is acceptable in streaming speech recognition.
We also downsample input representation with the convolution layers. As shown in Figure~\ref{fig:conv_transformer_encoder}, at the second convolution layer of each block, the stride is 2 along time dimension.
With an input frame rate of 10~ms, the frame rate becomes 20~ms in the first block, 40~ms in the second block, and finally 80~ms in the third block, as shown in Figure \ref{fig:delay-count}.
We find that this progressive downsampling scheme causes no loss in accuracy, and reduces memory requirement for training significantly. We can use bigger batch size for efficient training.
Moreover, downsampling greatly reduces computation cost of inference.
To further reduce computation complexity of audio encoder, we place most of Transformer layers in the third block with 80~ms frame rate.
The idea of using interleaved convolution for gradual downsampling and incorporating future context is inspired by \cite{peddinti2017low}.

\begin{figure}[th]
 %\vspace{-0.2cm}
  \centering
  \includegraphics[width=0.3\linewidth]{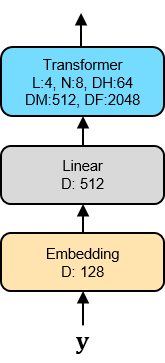}
  %\vspace{-0.3cm}
  \caption{Prediction net of Conv-Transformer Transducer. We denote the output dimensions of embedding layer and linear layer as \textbf{D}. }
  \label{fig:prediction_net}
\end{figure}

The prediction net is illustrated in Figure~\ref{fig:prediction_net}. An embedding layer converts previously predicted non-blank labels into vector representations. Then a linear layer projects the embedding vectors to match input dimension of unidirectional Transformer layers.

The joint net is a fully-connected feed-forward neural network with single hidden layer.
There are 512 units in the hidden layer, with Rectified Linear Unit (ReLU) as activation function.
We concatenate outputs of audio encoder and prediction net as input of joint net.

\subsection{Limiting history context of self-attention}

During decoding, self-attention of unidirectional Transformer attends to all history context. Computation cost continues to grow with the length of history context as decoding goes on, which is undesirable for a streaming system. We limit history context of self-attention with a fixed-size window. Computational cost of each step becomes constant.
The idea of limiting history context of self-attention for streamable speech recognition has been explored in \cite{wang2019transformer,yeh2019transformer,zhang2020transformer}.

\subsection{Relative position encoding}

We apply relative position encoding to model sequential order in self-attention \cite{dai2019transformer}.
Relative position encoding is proposed in \cite{shaw2018self}, and leads to significant improvement over absolute position encoding in the original Transformer.
Moreover, relative position encoding is necessary for reusing hidden states when applying self-attention with limited attention context \cite{dai2019transformer}.

\section{Experiments}

We perform our experiments with LibriSpeech dataset \cite{panayotov2015librispeech}, which is a publicly available English read speech corpus of audiobooks from the LibriVox project. The corpus consists of 960-hour training data, 10.7 hours of development data, and 10.5 hours of test data. The audio is sampled at 16~kHz and quantized at 16 bit. The development and test data are further divided into \texttt{clean} and \texttt{other} subsets according to the quality, assessed with a speech recognition system. We train our models with the entire training set. For clarity, we only report the results of test sets (test-clean and test-other), with the best-performed decoding configuration in the development sets.

\subsection{Training setup}

All the Conv-Transformer Transducer models are trained using 8 GPUs with per-GPU batch bucketing configuration as shown in Table~\ref{tab:batch_bucketing}. We use NovoGrad \cite{ginsburg2019stochastic} as our optimizer. The learning rate is warmed up from 0 to 0.01 in the first 10k steps, then is decayed to $5\times 10^{-6}$ in the following 200k steps polynomially.

We use 128-dimensional log-mel filterbank as acoustic feature, calculated with 20~ms window and 10~ms stride. We use 4k subwords as output target units, which are generated from training transcripts of LibriSpeech using SentencePiece \cite{kudo2018sentencepiece}.
We apply several techniques to avoid overfitting. We add additive Gaussian noise and apply speed perturbation \cite{ko2015audio} in time domain. After feature extraction, we apply SpecAugment as described in \cite{park2019specaugment}. We do not apply time-warping for SpecAugment as we have already applied speed perturbation. Transformer layers are regularized with 10\% dropout.
Due to limited computation resources, we choose hyperparameters without intensive tuning.

\begin{table}[th]
  \caption{Batch bucketing configuration}
  \label{tab:batch_bucketing}
  \centering
  \begin{tabular}{ c  c }
    \toprule
    \multicolumn{1}{c}{\textbf{SampleLength}} & 
                                         \multicolumn{1}{c}{\textbf{BatchSize}} \\
    \midrule
    $\ \ 0\sim\ \ 7s$     & $22$~~~            \\
    $\ \ 7\sim12s$        & $18$~~~            \\
    $12\sim14s$  &        $18$~~~              \\

    \bottomrule
  \end{tabular}
  \vspace{-5mm}
\end{table}

\subsection{Baseline conventional streamable model}

Our baseline system is based on Kaldi's \cite{povey2011kaldi} chain-model recipe \cite{povey2016purely}.
The TDNN-LSTM based acoustic model (AM) \cite{peddinti2017low} is trained with LibriSpeech, plus additional CommonVoice (CV) data (\texttt{en\_1488h\_2019-12-10}, validated set excluding dev and test sets) \cite{commonvoice}.
The AM is monophone-based, and contains 33.5M parameters. The look-ahead window size is 170~ms, which is comparable to our Conv-Transformer Transducer. We apply SpecAugment during AM training.
We perform language model (LM) training with normalized training text of the official LibriSpeech LM \cite{librispeechLM}, plus transcription of CV training data.
Vocabulary size of lexicon is about 200k. We deploy a two-pass decoding strategy, with a smaller 4-gram LM with 5.8M n-grams in first pass.
We utilize a bigger 4-gram LM with 191M n-grams for second-pass rescoring.
Note that we train the baseline system with additional CommonVoice data.

\subsection{Results}

We compare the results of our Conv-Transformer Transducer (ConvT-T) with the results of a previously published streamable Transformer Transducer model (T-T) \cite{zhang2020transformer}, a hybrid model with Transformer AM \cite{wang2019transformer} and our hybrid TDNN-LSTM AM baseline. We use beam search without an external language model for Conv-Transformer Transducer decoding. As shown in Table \ref{tab:wer_compare}, Conv-Transformer Transducer achieves the lowest WER on both test-clean and test-other. Our model operates with fewer parameters, smaller look-ahead window, and lower frame rate. We do not list the number of parameters of the hybrid models as they both rely on large external n-gram language models.

We also evaluate if low frame rate of 80~ms leads to performance degradation.
We train a model with 40~ms frame rate (HFR ConvT-T), by setting the stride to 1 instead of 2 in the second convolution layer of the third block.
As this model requires more memory, we train the 40-ms model with only half of the original batch size, but with twice more steps to maintain the same number of epochs. As shown in Table~\ref{tab:wer_compare}, the performance difference between the 40-ms model and the 80-ms model is insignificant.

\begin{table}[th]
  \centering
  \caption{WER comparison with previously published streamable Transformer Transducer model, hybrid streamable model with a Transformer-based AM, our conventional hybrid baseline model and our streamable Conv-Transformer Transducer models on LibriSpeech test sets.}
  \scalebox{0.8}{
  \begin{tabular}{c|c|c|c|c|c}
  \toprule
  \multirow{2}*{Model} & \multirow{2}*{Params} & Look & Frame & \multicolumn{2}{c}{WER (\%)} \\
   & & ahead & rate & clean & other \\
  \midrule

  Hybrid TDNN-LSTM & - & 170ms & 30ms & 4.0 & 8.9 \\ 
  Hybrid Transformer {\cite{wang2019transformer}} & - & 2480ms & 20ms & 3.65 & 9.01 \\
   T-T {\cite{zhang2020transformer}} & 139M & 1080ms & 30ms & 3.6 & 10.0 \\
   ConvT-T (Ours) & 67M & 140ms & \textbf{80ms} & \textbf{3.5} & 8.3 \\
   HFR ConvT-T (Ours) & 67M & 140ms & 40ms & \textbf{3.5} & \textbf{8.2} \\
  \bottomrule
  \end{tabular}
  }
  \label{tab:wer_compare}
  %\vspace{-5mm}
\end{table}

We then perform experiments to analyze effects of varying the number of Transformer layers over different blocks of various frame rates.
The results are shown in Table \ref{tab:layer_distribution}. Recall that in the original model, there are 2 layers of Transformer in the first (20~ms) block, 2 layers of Transformer in the second (40~ms) block and 8 layers of Transformer in the third (80~ms) block. Moving Transformer layers from the third to the second block does not affect recognition performance significantly, but increases computation cost as more Transformer layers operate at higher frame rate.
However, when we move all the Transformer layers in the second block to the third block as shown in Row 4 of Table \ref{tab:layer_distribution}, small performance degradation begins. We further move all the Transformer layers in the first two blocks to the third block as shown in Row 5. This is equivalent to running a convolutional neural network (CNN) followed by multi-layer Transformer. There is further performance degradation. Note that there are fewer parameters for the Transformer layers in the first block. We reduce total number of Transformer layers from 12 to 11 to maintain similar number of parameters. The experimental results support our strategies of placing most of Transformer layers in the third block with the lowest frame rate, and applying interleaved convolution in the model.

\begin{table}[th]
 \centering
 \caption{WER of different number of Transformer layers over the three blocks in audio encoder. Layer distribution B1-B2-B3 corresponds to number of Transformer layers in the first, the second and the third block respectively}
 \begin{tabular}{c|c|c}
 \toprule
 Layer & \multicolumn{2}{c}{WER (\%)} \\
 distribution & clean & other \\
 \midrule
  2-2-8 & 3.5 & \textbf{8.3}  \\
  2-4-6 & 3.5 & \textbf{8.3}  \\
  2-6-4 & \textbf{3.4} & 8.4  \\
  2-0-10 & 3.6 & 8.5 \\
  0-0-11 & 3.7 & 8.9 \\
 \bottomrule
 \end{tabular}
 \label{tab:layer_distribution}
 %\vspace{-5mm}
\end{table}

We further evaluate the effects of limiting the amount of history context (left-context) in self-attention of Transformer layers.
The results are shown in Table \ref{tab:limited_attention}.
The size of attention window in prediction net does not have a significant impact on recognition performance of test-clean, even when the window size is only 2. With an attention window size of 16, we get the same accuracy as the original model.
A bigger window helps to improve accuracy of the more challenging test-other slightly. For the evaluation of audio encoder, we fix the window size of self-attention to 16 in prediction net. Unlike prediction net, a smaller window in self-attention of audio encoder leads to more significant performance degradation. With self-attention window size of 32 in audio encoder, although there is performance degradation, overall results are still competitive to those of hybrid models and Transformer Transducer.

\begin{table}[th]
  \centering
  \caption{WER of different window sizes of history context (left-context) in self-attention of Transformer layers. Inf means that all history context is used.}
  \begin{tabular}{c|c|c|c}
  \toprule
  \multicolumn{2}{c|}{Left-context window size} & \multicolumn{2}{c}{WER (\%)} \\
   audio encoder & prediction net & clean  & other \\
  \midrule
   Inf & Inf & 3.5 & \textbf{8.3} \\
   Inf & 2   & 3.5 & 8.6 \\
   Inf & 8   & 3.6 & 8.5 \\
   Inf & 16  & \textbf{3.4} & \textbf{8.3} \\
   8   & 16  & 3.8 & 9.3 \\
   16  & 16  & 3.6 & 9.1 \\
   32  & 16  & 3.6 & 8.9 \\
  \bottomrule
  \end{tabular}
  \label{tab:limited_attention}
  %\vspace{-5mm}
\end{table}

\section{Conclusions}

We propose Conv-Transformer Transducer model, which is suitable for streaming speech recognition.
Our model is end-to-end trainable using Transducer framework.
By applying unidirectional Transformer with interleaved convolution for audio encoding, our model requires only a 140~ms look-ahead window. A smaller look-ahead window generally leads to lower latency. We gradually downsample audio input into frame rate of 80~ms to reduce computation cost. Recognition accuracy is still maintained. Experimental results also support our strategies of applying interleaved convolution, and placing most of Transformer layers in the block with the lowest frame rate. We also limit the size of history context window in self-attention of Transformer layers. Although there is performance degradation, the results are still comparable to published LibriSpeech results.  Our model achieves competitive results to previously published streamable Transformer transducer model, and state-of-the-art hybrid models, with lower latency, lower frame rate and fewer parameters.

\bibliographystyle{IEEEtran}

\bibliography{mybib}

\end{document}